# AESSI: An Around-Ear Silent Speech Interface for Cross-Day Online Reuse without Test-Day Calibration

Xiran Xu, Mochu Dong, Yujie Yan, Chenxi Wang, Yu Jiao, Jing Chen*

Silent speech interfaces (SSIs) enable private communication without audible speech and may support people with post-stroke dysarthria. Everyday reuse requires articulation-related representations that generalize across days despite sensor repositioning and physiological changes. We present AESSI, an around-ear SSI using masked-context representation pretraining (MCRP): a student predicts teacher representations from masked time-frequency inputs to encourage robustness to recording variability. We collected 44 electrophysiological recordings from 24 participants for 25 everyday Mandarin sentences. With six participants' complete final recordings held out, AESSI achieved 92.24% mean accuracy without test-day calibration. AESSI exceeded the best adapted baseline by 50.57 percentage points. At least 21 days after each participant's last recording, five participants each completed 50 independently randomized online tests without calibration, achieving 98.0% overall accuracy. Median preprocessing and inference time in CPU replay was 31.70 ms. These results demonstrate end-to-end system operation and support online reuse without test-day calibration. Demo is with the paper.

## 1 INTRODUCTION

Speech provides an efficient, hands-free and eyes-free channel for interacting with computing devices [1, 9]. Speaking aloud, however, is not appropriate in every situation: speech in public spaces can disturb others, disclose private information, and be affected by environmental noise [8, 20]. People with post-stroke dysarthria, those who have undergone total laryngectomy, and people with incomplete locked-in syndrome may also have difficulty consistently producing intelligible audible speech [14, 25, 30]. Silent speech interfaces (SSIs) sense activity associated with silent articulation and decode the linguistic content that a user intends to express [32]. Previous work has captured lip movements from facial images [33] and observed tongue movements with submental ultrasound [20]. Among electrophysiological approaches, AlterEgo uses neuromuscular signals from the face and neck for personalized wearable silent interaction [18], while Gaddy and Klein [11] convert electromyographic signals recorded during silent articulation into audible speech. Other studies use full-scalp electroencephalography (EEG) or joint EEG–electromyography (EMG) recordings to capture electrophysiological signals associated with silent speech [15, 16].

Bilateral around-ear electrode arrays [5] offer a promising setting for investigating this problem. Electrodes on the skin around the ears leave the oral cavity unobstructed and could be integrated into ear-worn devices, making the layout attractive for everyday wear. Prior work has demonstrated that around-ear electrophysiological signals can support online silent speech decoding [14]. We focus on this interaction setting (Figure 1a). For a sensing layout that depends on skin contact, an important use scenario is that a user puts the device back on at a later date and receives reliable online recognition after each silent expression, without collecting new labeled calibration data. We refer to this scenario as cross-day online decoding without test-day calibration. The central question is how to learn representations that continue to distinguish silent expressions despite changes in wearing conditions and recording dates, so that a historically trained model remains usable after the electrodes are reapplied on later days.

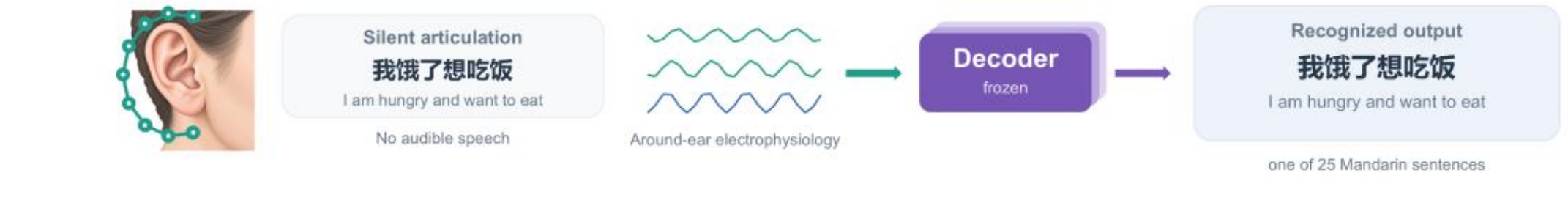

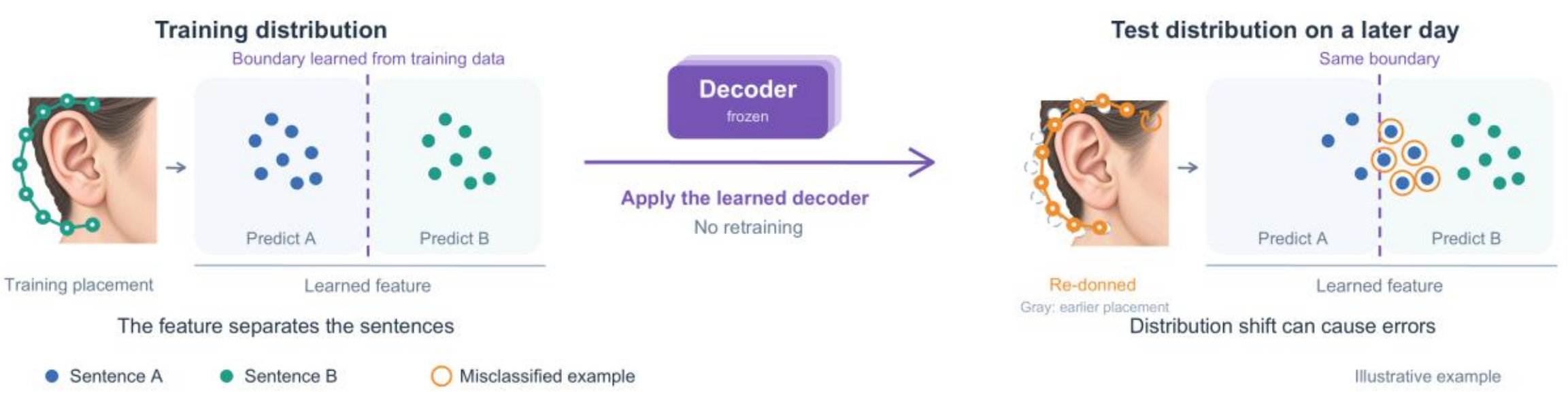

Figure 1: Around-ear silent speech interaction across days. (a) Bilateral around-ear electrode arrays record electrophysiological activity during silent articulation. After a 3 s observation window, a frozen decoder outputs one of 25 predefined Mandarin sentences. (b) Reapplying electrodes on a later day may change the representation of the same expression, challenging reuse of a historically trained model. Electrode locations, waveforms, representation distributions, and sentence outputs are schematic. The example sentence is 'I am hungry and want to eat.'

Two related challenges arise. First, electrophysiological observations depend not only on linguistic content but also on electrode placement and contact, the user's physiological state, and articulation strategies [5, 17, 29]. Reapplying electrodes on a later day can therefore shift the signal distribution between historical training and future testing (Figure 1b). Without test-day calibration, a model must learn articulation-related structure from historical recordings that remains discriminative after reapplication, reducing dependence on a particular wearing configuration. Second, online decoding requires timely feedback after each silent expression. Processing and recognition must use only the short signal window already available, without using labeled or unlabeled test-day data outside that window.

Wearable electrode-array SSIs have advanced through facial and neck surface EMG (sEMG), joint EEG–EMG recordings, and around-ear arrays [14, 15, 24, 28, 29, 31, 34]. For cross-day reuse, Gaddy [10] studied EMG decoding across recording sessions, and SilentSpeller used a custom electropalatography device for live online text entry approximately one week after the most recent phrase-training data were collected [19]. These studies show that historical training can support subsequent use, although reuse must be evaluated for each sensing layout and reapplication condition. For example, recognition accuracy in the neck-sEMG system SilentWear declined substantially when the neckband was worn again on different days; fine-tuning with additional data from the new day improved performance [29]. Prior work has also explored several routes to online decoding. Gaddy [10] developed a real-time streaming speech prototype; SilentWear deployed its decoding network on a microcontroller to demonstrate low-latency wearable inference [29]; and MindSpeak integrated acquisition, decoding, and text feedback into a real-time silent interaction system [37]. For around-ear arrays, Inoue et al. [14] demonstrated online decoding of 64 words, but healthy participants first completed 640 labeled trials on the test day, which were combined with historical recordings for fine-tuning. This preparation adds to the effort required before a user can interact after reapplying the device. Maintaining recognition while reducing repeated calibration remains an obstacle to everyday reuse of skin-contact around-ear electrodes. Prior work has thus advanced both cross-day decoding and real-time interaction; around-ear SSIs still require cross-day online decoding without test-day calibration so that users can directly reuse their historical models after putting the device on again.

We present AESSI (Around Ear Silent Speech Interface), an online SSI that uses bilateral around-ear electrophysiological signals to recognize 25 everyday Mandarin sentences. Compared with isolated words and control commands, sentence-level targets convey more complete interaction intents. A 25-class task provides a manageable balance between coverage of everyday expressions, repeated cross-day collection, and the number of examples per class, enabling systematic evaluation of model reuse after electrode reapplication. To support online decoding, AESSI converts each 3 s silent-articulation window into a log short-time Fourier transform (log-STFT), encodes time-frequency and spatial features through four branches, and aggregates class-related information using temporal convolutions and attention to produce one prediction at the end of the window (Figure 4). Preprocessing and decoding use only data available by that time; all fitted parameters are determined from source training sessions. To address changes caused by electrode reapplication and day-specific conditions, AESSI combines source-side augmentation with single-teacher masked-context representation pretraining (MCRP), aiming to reduce dependence on particular recording conditions or a small set of local cues. In MCRP, a frozen teacher produces target representations from complete inputs, and a student predicts these representations from masked inputs, encouraging the use of context distributed across time and frequency to learn class-related structure.

We collected 44 historical sessions from 24 participants and evaluated cross-day decoding on later-day recordings from six participants with multiple sessions. AESSI achieved a participant-macro-averaged accuracy of 92.24% without test-day calibration (chance: 4.00%). Five of these participants subsequently returned at least 21 days after their most recent historical recording, had the electrodes reapplied, and completed 50 trials each in a prospective live online evaluation. The four-branch decoding model and processing pipeline were frozen before evaluation. Targets were generated on site through RANDOM.ORG on an independent tablet, and no test-day model adaptation was performed. The system correctly recognized 245 of 250 trials, yielding 98.00% online accuracy. All five evaluations were continuously recorded from electrode placement through preparation, articulation, and visible feedback. One representative evaluation video is provided with the paper. All five videos will be made public after double-blind review.

This paper makes the following contributions:

1. We conduct an around-ear silent speech study comprising 24 participants and 44 historical sessions, and evaluate cross-day model reuse supported jointly by personal history and population data using later-day sessions from six participants.
2. We introduce a decoding method combining an online-compatible time-frequency representation, four-branch encoding, source-side augmentation, and MCRP. It learns class-related structure through contextual prediction under partial masking and achieves 92.24% mean cross-day accuracy across three random seeds on the 25-class task.
3. We implement and prospectively validate online reuse of AESSI without test-day calibration. After at least 21 days, five participants have their around-ear electrodes reapplied and directly use frozen models for 250 live trials, achieving 98.00% accuracy. Replay of the raw trials on the deployment CPU yields a median preprocessing and sentence-prediction time of 31.70 ms.

## 2 RELATED WORK

### 2.1 Silent Speech Interfaces and Around-Ear Electrode Arrays

SSI sensing designs must balance access to articulation information, the space occupied by the device, and everyday wearing requirements [32]. Vision-based approaches recognize silent expressions from lip and jaw movements and require the mouth to remain within the sensor's field of view [33]. Submental ultrasound observes tongue movements using a probe beneath the chin [20]. Electropalatography records tongue–palate contact through a custom intraoral device and has supported mobile silent text entry [19]. Full-scalp EEG provides broad electrode coverage, typically requiring a cap and wiring across hair-bearing regions [5, 15, 16]. These approaches access different articulation or electrophysiological information and impose different wearing requirements.

Facial and neck EMG form another major line of SSI research. AlterEgo demonstrated personalized wearable silent interaction using neuromuscular signals [18]. Gaddy and Klein [11] aligned voiced and silent articulation recordings to transfer audio supervision to silent EMG, enabling conversion of silent articulation into audible speech. Subsequent work encoded EMG directly with convolutions and a Transformer and introduced auxiliary phoneme supervision to improve synthesized speech intelligibility [12]. These studies established a technical basis for EMG-driven silent interaction and speech generation, while highlighting electrode layout and stability across recordings as system-design considerations.

Noninvasive SSIs have explored different scales of linguistic output. Systems such as AlterEgo primarily recognize digits, commands, or predefined phrases, and recent wearable systems also commonly use limited vocabularies for online validation [14, 18, 31]. Meanwhile, facial-EMG speech synthesis and electropalatographic spelling have begun to support large-vocabulary input [11, 12, 19]. These advances provide a basis for moving from isolated commands toward sentence-level everyday interaction and motivate our use of 25 everyday sentences to evaluate cross-day reuse.

Around-ear arrays provide an alternative electrode location for electrophysiological SSIs. Electrodes can be placed on relatively hair-free skin around the ears, leave the oral cavity unobstructed, and potentially integrate with headphones, earmuffs, or hearing devices [5, 7, 31]. Ear-centered electrophysiology has already demonstrated prolonged mobile recording and integration with functional hearing devices [6, 7]. Everyday SSI reuse additionally requires evaluating decoding after reapplication and the user experience of specific systems.

Around-ear electrodes lie near temporal, jaw, and facial muscles. Biting, chewing, speaking, and facial movements produce detectable electrophysiological changes at these locations, providing a basis for sensing silent articulation [14, 17, 21]. Electrode position and spatial coverage also affect what can be recorded. Ear-centered auditory-attention research shows that different layouts influence target-speaker detection [26]. Silent-articulation information in around-ear recordings therefore needs direct task-specific evaluation, with decodability and cross-day reuse assessed separately.

Inoue et al. [14] used bilateral around-ear arrays for 64-word silent speech recognition, combining heterogeneous training data from around-ear electrodes, full-scalp EEG, and facial EMG. They demonstrated online interaction in healthy participants and one participant with incomplete locked-in syndrome. This directly established the feasibility of practical silent interaction using around-ear signals. Their online procedure included personal calibration on the test day: healthy participants completed 10 rounds of 64 words, which were combined with historical recordings to fine-tune the model. The participant with incomplete locked-in syndrome completed 10, 12, and 14 calibration rounds before the three online tests, respectively. This procedure supplies data for adaptation to that day's signals but increases preparation before each interaction session. We investigate how existing personal and population recordings can instead support direct reuse of an around-ear model after electrode reapplication on a future day.

### 2.2 Cross-Day Generalization and Online Reuse of Silent Speech Interfaces

Evaluating everyday SSI reuse requires distinguishing recognition under the same recording conditions, cross-day generalization, and actual online use. We define a session as one complete continuous recording after a single electrode application. Most experiments, including our test evaluations, contain only one session per day; in that case, cross-day and cross-session generalization coincide. In a leave-one-session-out split, an entire session is reserved for testing and the remaining sessions are used for training (Figure 2a). By contrast, some studies use within-session splits [15, 24, 28, 34], in which training and testing share an electrode application and similar acquisition conditions. Temporal autocorrelation in physiological signals such as EEG, or recording characteristics that covary with the task, can inflate decoding accuracy under such splits. Within-session performance alone does not establish performance on a later day [23, 35].

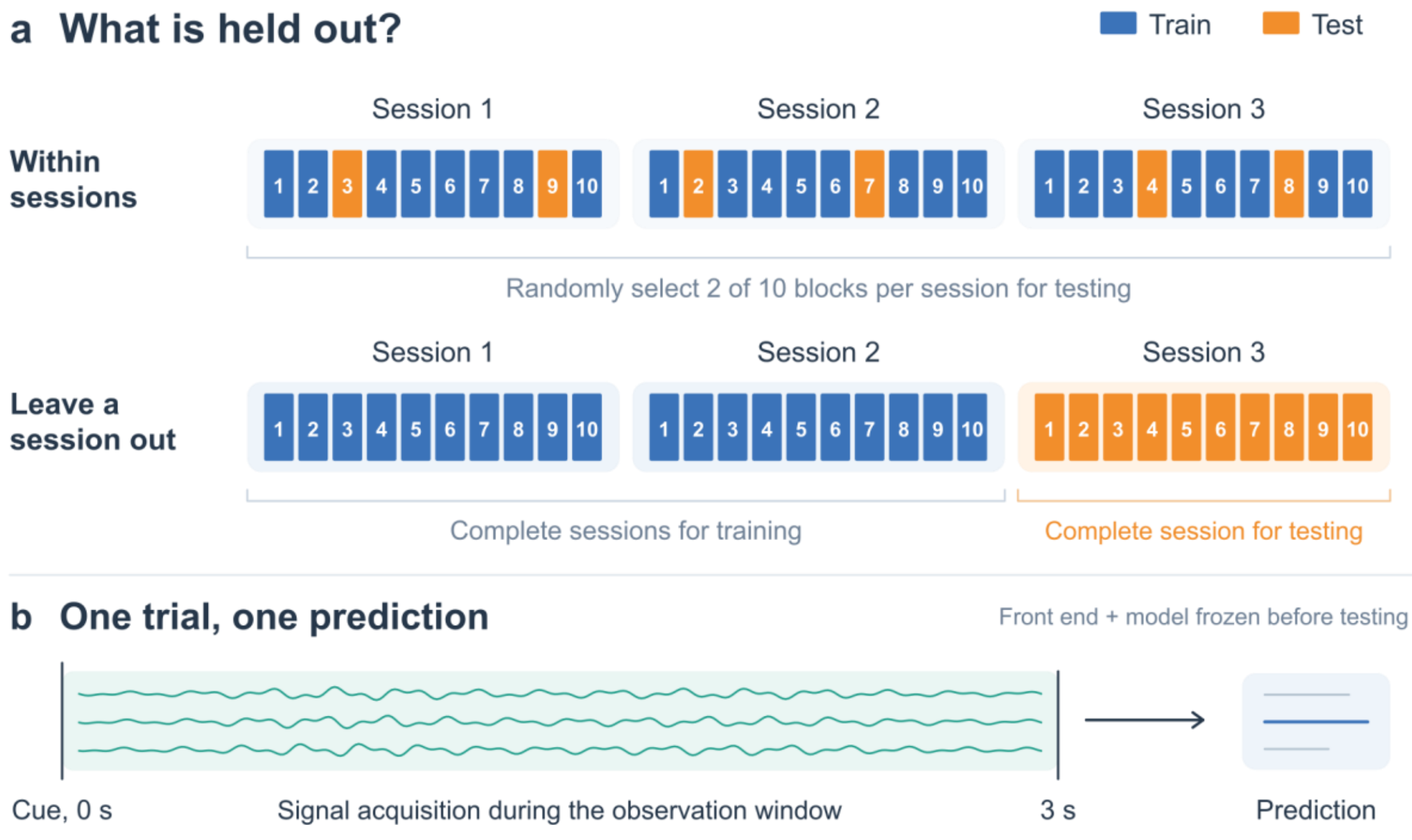

Figure 2: Study design and boundaries of future-session evaluation. (a) Within-session splits share recording conditions between training and testing; session-level holdout reserves an entire session exclusively for testing. Blocks are schematic, and the validation split within source data is omitted. (b) Each cue is followed by a fixed 0–3 s observation window. One prediction uses only signals available by the end of that window. Preprocessing parameters and the model are fixed before the target session begins. Waveforms and temporal spacing are schematic.

Prior work has contributed to cross-session decoding and online reuse after historical training. Gaddy [10] investigated EMG decoding with entire sessions held out and developed a real-time streaming speech prototype. SilentSpeller used a custom electropalatography device for live online text entry approximately one week after the latest phrase-training recording, providing direct evidence of reuse across days [19]. SilentWear evaluated cross-day generalization of neck-sEMG decoding through repeated wearing on different days and leave-one-session-out evaluation, and studied incremental fine-tuning with new-session data. Its decoding network was also deployed on a microcontroller to evaluate on-device computational performance [29]. MindSpeak combined holdout evaluation separated by date with a real-time interaction prototype, further connecting cross-day decoding with online implementation [37]. Building on these advances, we make the absence of test-day calibration a central goal for everyday reuse of around-ear SSIs: historically learned recognition should remain available after reapplication, allowing users to begin silent interaction without supplying another calibration corpus or adapting the model. Our online evaluation starts with electrode placement and tests whether the existing system directly supports silent expression and real-time feedback after reapplication on the test day.

## 3 METHOD

### 3.1 Problem Formulation and System Objectives

We study around-ear silent sentence recognition from a fixed candidate set, treating each independently donned recording session as a domain. Each example consists of a bilateral around-ear electrophysiological observation $x$ and its sentence class $y$, where $y$ denotes one of 25 target sentences. Sessions recorded from the same user on different days belong to different domains.

Following the treatment of domain covariates in ManyDG [36], we use a latent variable $z$ to represent recording conditions shared within a session, including electrode placement, contact, and the user's physiological state that day. The observation $x$ depends jointly on sentence class $y$ and session covariate $z$, with the generative process written as

$$z \sim p(z), \quad x \sim p(x \mid z, y).$$

Here, $z$ describes shared session-level recording conditions, while the conditional distribution $p(x \mid z, y)$ captures trial-level differences in movement and noise. Reapplication and use across days may change $z$, producing different signal distributions for the same sentence across sessions (Figure 1b). Good performance on observed sessions therefore does not imply that a model can be used directly in a new session. Randomly splitting examples within a session also does not adequately test cross-session generalization (Figure 2a). Moreover, the system must return an online result after acquiring a short window of silent-speech electrophysiology. It therefore cannot use additional unlabeled data from the test session outside that window to fit the test-day distribution. Our task is consequently one of domain generalization rather than domain adaptation.

Our main objective is to learn a mapping $f: x \rightarrow y$ from population recordings of other users and the target user's historical sessions that continues to recognize sentences accurately in that user's new session. AESSI thus aims to retain class information in $x$ that transfers across sessions while reducing dependence on recording features associated with session covariate $z$.

### 3.2 Participants, Equipment, and Data Collection

Twenty-four native Mandarin speakers participated (11 women; age range: 18–46 years). All were right-handed, reported no hearing or speech impairments, and had no history of brain injury or cognitive deficits. The protocol was approved by the relevant biomedical ethics committee; identifying details will be disclosed after double-blind review. All participants provided written informed consent before the experiment. Sessions took place in an EEG-shielded room or an open meeting room. Each session lasted approximately 40 minutes, and participants received approximately US$15 per session.

Signals were acquired at 1 kHz with a TMSi SAGA 32+ amplifier and bilateral cEEGrid around-ear arrays comprising 20 electrodes in total (Figure 3a; [5]). An additional electrode on the participant's left arm served as ground, and an average reference was used. Channels 18 and 20 were consistently excluded from recording because of high contact impedance, leaving 18 electrophysiological input channels. Before each experiment, all 18 channels had impedances below 20 kΩ and no remaining bad channels. No online filtering was applied during signal recording. Each independent electrode application produced one session of 10 blocks. Within each block, all 25 sentences appeared once in random order, yielding 250 trials per session. Across 24 participants, we collected 44 sessions and 11,000 trials. Six participants completed at least three sessions and served as the target users for cross-day follow-up evaluation.

Each trial presented the target sentence for 2 s, a preparation cue for 0.5 s, a silent-expression cue for 3 s, and a 1 s rest interval (Figure 3b). The program recorded the onset trigger for the silent-expression stage. Participants were instructed to silently articulate the target sentence once during the 3 s cue and to try to avoid whispering. Only this 3 s stage was used as the decoding window.

**a Recording hardware**

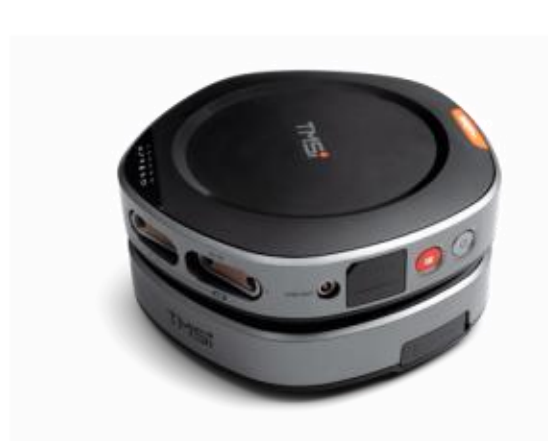

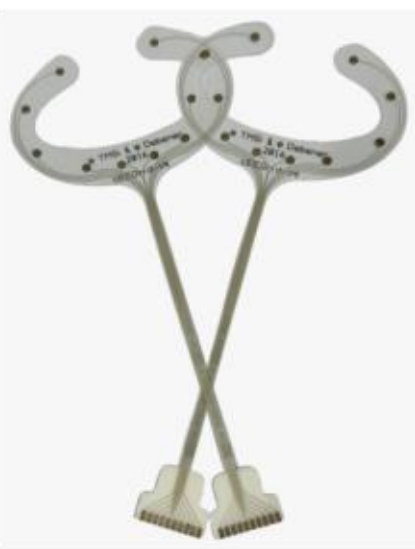

**b Trial sequence**

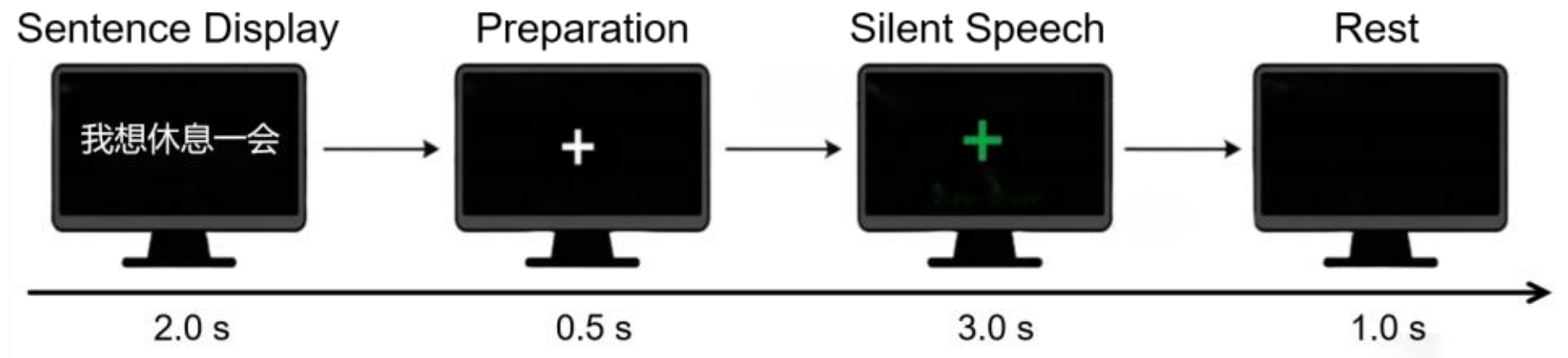

Figure 3: Around-ear acquisition equipment and trial sequence. (a) The TMSi SAGA 32+ amplifier and bilateral cEEGrid arrays. (b) A trial consists of target presentation, preparation, silent articulation, and rest. The model uses the 3 s silent-articulation stage. The Mandarin example on the display means 'I want to rest for a while.'

One potential application is short-sentence communication support for people with post-stroke dysarthria. Drawing on the design of vocabularies for everyday communication [27], we selected 25 short Mandarin sentences covering daily activity needs, social interaction, and physical and emotional states (Table 1). Each sentence contains six Chinese characters, reducing obvious class cues from sentence length. Table 2 lists the number of sessions and collection settings for each participant.

Table 1: The 25 experimental sentences, using original IDs 01–25. Each Mandarin sentence is followed by the English translation used in the online demonstrations.

| Daily activity needs | Social interaction | Physical and emotional states |
|---|---|---|
| 01. 我需要去厕所<br>I need to go to the restroom | 16. 你感觉怎么样<br>How do you feel? | 11. 我感觉很舒适<br>I feel very comfortable |
| 02. 我想休息一会<br>I want to rest for a while | 17. 你打算去哪里<br>Where are you planning to go? | 12. 我感觉不舒服<br>I feel uncomfortable |

| Daily activity needs | Social interaction | Physical and emotional states |
|---|---|---|
| 03. 帮我打个电话<br>Please make a phone call for me | 18. 很高兴见到你<br>Nice to meet you | 13. 我心情很开心<br>I am very happy |
| 04. 把我眼镜拿来<br>Bring me my glasses | 19. 谢谢你的帮助<br>Thank you for your help | 14. 我心情很难过<br>I feel very sad |
| 05. 我想出去走走<br>I want to go out for a walk | 20. 请你稍等一会<br>Please wait a moment | 15. 我感觉很疲惫<br>I feel very tired |
| 06. 我需要换衣服<br>I need to change clothes | 21. 我很喜欢这个<br>I really like this | |
| 07. 请给我量体温<br>Please take my temperature | 22. 你正在做什么<br>What are you doing? | |
| 08. 请叫医生过来<br>Please call the doctor | 23. 请你再说一遍<br>Please say that again | |
| 09. 我渴了想喝水<br>I am thirsty and want some water | 24. 今天天气怎样<br>How is the weather today? | |
| 10. 我饿了想吃饭<br>I am hungry and want to eat | 25. 你的心情怎样<br>How are you feeling? | |

Table 2: Number of sessions and data collection settings for each participant.

| Participant | Sessions | EEG-shielded room / Meeting room | Participant | Sessions | EEG-shielded room / Meeting room |
|---|---|---|---|---|---|
| P001 | 1 | 1 / 0 | P013 | 1 | 1 / 0 |
| P002 | 3 | 1 / 2 | P014 | 1 | 1 / 0 |
| P003 | 10 | 5 / 5 | P015 | 3 | 1 / 2 |
| P004 | 3 | 1 / 2 | P016 | 1 | 1 / 0 |
| P005 | 1 | 1 / 0 | P017 | 1 | 1 / 0 |
| P006 | 1 | 1 / 0 | P018 | 4 | 1 / 3 |
| P007 | 1 | 1 / 0 | P019 | 1 | 1 / 0 |
| P008 | 1 | 1 / 0 | P020 | 1 | 1 / 0 |
| P009 | 1 | 1 / 0 | P021 | 1 | 1 / 0 |
| P010 | 1 | 1 / 0 | P022 | 1 | 0 / 1 |
| P011 | 1 | 1 / 0 | P023 | 1 | 1 / 0 |
| P012 | 1 | 1 / 0 | P024 | 3 | 1 / 2 |

Setting columns give the number of sessions in each location. The 44 sessions comprised 27 in the EEG-shielded room and 17 in the meeting room. Only anonymous participant IDs are reported.

### 3.3 Log-STFT Input Representation

Online decoding must produce a result using only the short window available at prediction time (3 s). Neither preprocessing nor decoding can depend on future samples or statistics from the complete test dataset, and excessive computational complexity would introduce undesirable latency. To avoid discrepancies caused by different training and test preprocessing,

AESSI applies the same log-STFT representation in both stages, retaining spectral structure and temporal changes during articulation. Specifically, for each fixed 3 s, 18-channel recording sampled at 1 kHz, we compute a one-sided STFT with a 200 ms Hann window and a 50 ms hop, after prepending 200 zeros. We retain 100 frequency bins from 5 to 500 Hz at 5 Hz intervals and apply logarithmic compression with a fixed floor:

$$U = 10\log_{10}\left(\max(|\mathrm{STFT}(x)|^2, 10^{-12})\right).$$

The maximum is taken elementwise. The resulting input U has shape 18 × 100 × 61, corresponding to 18 electrode channels, 100 frequency bins, and 61 time frames.

### 3.4 Four-Branch Temporal Decoder

The decoder extracts four representations from the same input and fuses them along a shared temporal axis (Figure 4a). The time-frequency branch retains local spectral changes over time, applying two 2D convolutions followed by averaging over frequency. The frequency-aggregation branch first averages across electrodes and then uses 1D temporal convolutions to encode the trajectory of each frequency bin. The channel branch first averages across frequencies to encode each electrode's trajectory of mean spectral intensity:

$$H_{\mathrm{TF}} = \mathrm{Mean}_f(\mathrm{TFConv}(U)),$$
$$H_{\mathrm{Freq}} = \mathrm{FreqConv}(\mathrm{Mean}_c(U)),$$
$$H_{\mathrm{Channel}} = \mathrm{ChannelConv}(\mathrm{Mean}_f(U)).$$

Subscripts on Mean indicate the dimension being averaged; the temporal axis is preserved. ChannelConv includes a learnable linear gate at its output. The Gram branch captures relationships between the spectral shapes of different electrodes at the same time point. Each channel's spectrum in the current frame is centered and normalized to unit length, forming matrix $\overline{U}_t$. We then construct a shrinkage Gram matrix:

$$G_t = 0.95\overline{U}_t\overline{U}_t^{\mathrm{T}} + 0.05I_{18}, H_{\mathrm{Gram}} = \mathrm{GramConv}(\mathrm{svec}(G)).$$

For each frame, svec extracts the 171 upper-triangular elements, including the diagonal, and multiplies off-diagonal elements by √2. GramConv combines pointwise projection, framewise normalization, and depthwise temporal convolution into a residual representation, followed by a linear gate. Each branch outputs a 96 × 61 sequence. The four sequences are concatenated along the feature dimension and fused into a 192 × 61 sequence using a 1 × 1 convolution, group normalization, and SiLU:

$$H_{\mathrm{Fusion}} = \mathrm{Fuse}([H_{\mathrm{TF}}, H_{\mathrm{Freq}}, H_{\mathrm{Channel}}, H_{\mathrm{Gram}}]).$$

The fused sequence passes through five temporal residual blocks, each combining depthwise temporal convolution, pointwise transformations, and channel recalibration. Four-head self-attention and a feedforward layer then produce the window representation H. Temporal convolutions use left zero-padding; normalization, channel recalibration, and attention operate within the complete, already acquired observation window. Pooling computes an attention-weighted mean, an ordinary mean, a maximum, and a standard deviation, concatenating them into a 768-dimensional vector. A 160-dimensional projection head and a 25-class linear classifier produce the output probabilities:

$$p = \mathrm{Softmax}\left(\mathrm{FC}_{25}(\mathrm{Head}_{160}(\mathrm{Pool}(H)))\right).$$

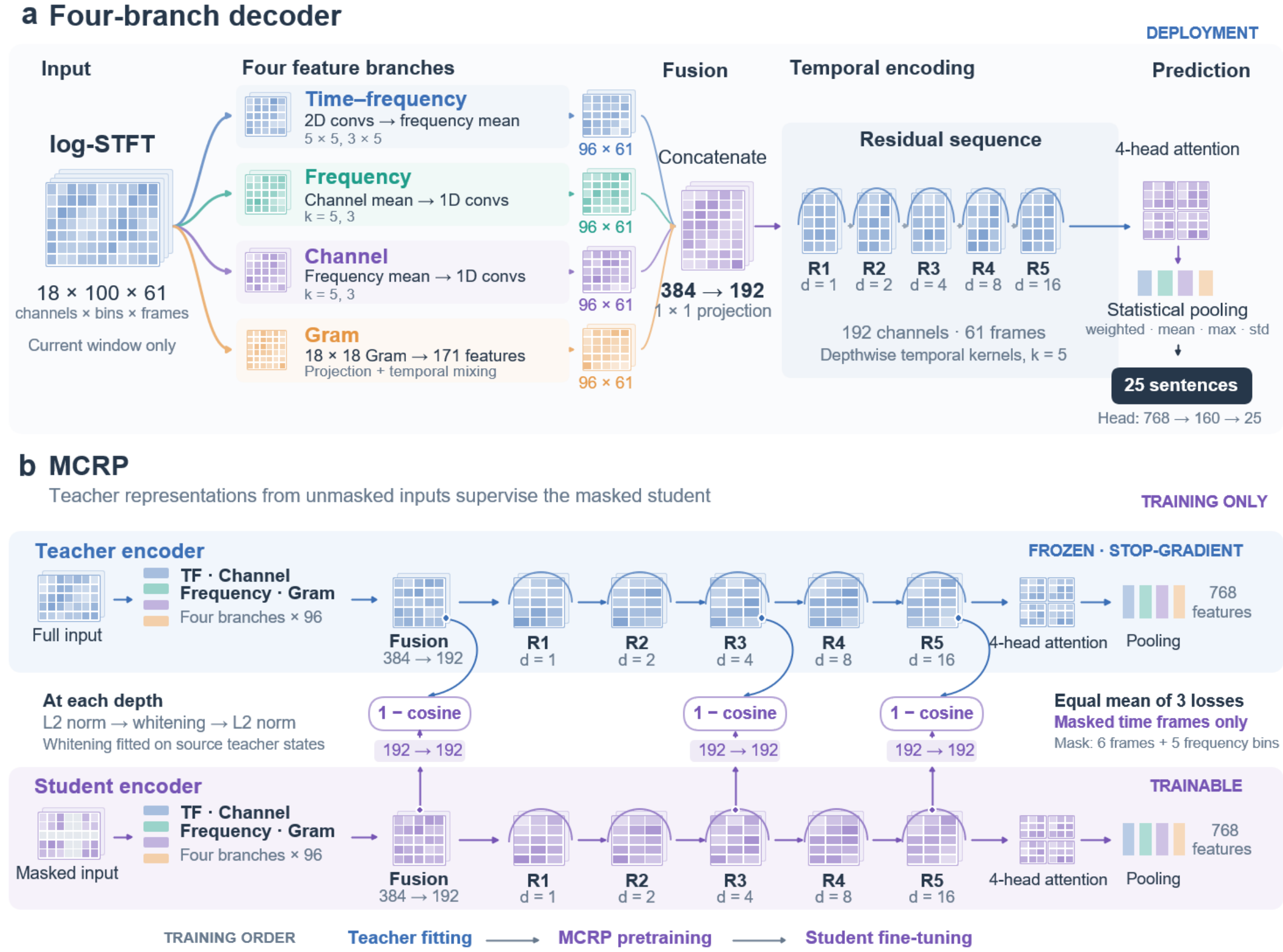

Figure 4: Four-branch decoder and single-teacher masked-context representation pretraining. (a) Four branches extract and fuse representations from the same log-STFT input to predict one of 25 sentences. (b) Teacher and student share the same decoder architecture. The teacher is frozen after supervised training. With portions of its time-frequency input masked, the student learns to align its intermediate representations at Fusion, R3, and R5 with those of the teacher, followed by fine-tuning. Only the student is deployed in AESSI.

### 3.5 Data Augmentation and Masked-Context Representation Pretraining

Cross-session decoding requires representations to retain information about sentence class y as covariate z changes. Electrode reapplication and day-specific physiological changes can alter the reliability of local time-frequency cues, weakening or invalidating features that were useful in training sessions. We therefore combine data augmentation with MCRP (Figure 4b). Controlled time-frequency masking approximates disruptions to some local cues and encourages the model to maintain class-related representations using the remaining context.

During sentence-supervised training, we apply random augmentation (Aug) to log-STFT tensors on the fly without changing their sentence labels. Each mini-batch selects one transformation from a predefined mixture: no augmentation, temporal shifting alone, temporal shifting followed by temporal masking, temporal shifting followed by channel perturbation, frequency masking, or spectral tilting. Their selection probabilities are 10%, 50%, 10%, 10%, 10%, and 10%, respectively. The strategy is selected per batch, and its parameters are resampled on each invocation.

For strategies involving temporal shifting, an integer displacement is sampled uniformly from −8 to +8 frames and shared across the batch. Features shifted outside the window are discarded, and newly exposed positions are zero-filled. Temporal and frequency masking cover 1–5 consecutive time frames or frequency bins, respectively. Mask width is sampled from a discrete uniform distribution, and the starting position is sampled uniformly from positions that accommodate the complete mask. Width and position are shared across the batch. Temporal masking spans all channels and frequencies, while frequency masking spans all channels and time frames; masked features are set to zero.

For channel perturbation, each channel of each example is independently zeroed across all time-frequency features with probability 0.1. Retained channels are multiplied by Gaussian random scale factors with mean 1 and standard deviation 0.08, constant across the time and frequency dimensions of that channel. Spectral tilting independently samples a slope a for each example from the uniform distribution on [−0.2, 0.2] and multiplies frequency bin f by $1 + a \cdot u_f$, where $u_f$ is its normalized frequency coordinate, evenly spaced from −1 to 1. Each example's tilt is shared across all channels and time frames. These augmentations are used only during supervised training; validation and testing use no random augmentation.

MCRP draws on masked reconstruction learning in vision [13] and self-distillation that predicts contextual representations of complete inputs from masked inputs [3]. We use a teacher frozen after supervised training and train a student to predict its intermediate representations of complete inputs. All teacher parameters remain frozen, and the student is initialized from scratch. Teacher and student have identical architectures. For each student input, we randomly zero six consecutive time frames across all frequencies and five consecutive frequency bins across all time frames. At the fusion layer and the outputs of residual blocks 3 and 5, separate 192-to-192 projections, implemented as randomly initialized 1D convolutional layers, align student representations with the teacher's native 192-dimensional representations.

To balance scales across feature directions, we estimate the mean and whitening matrix of teacher representations at each layer using source training examples. Both teacher and student representations are first normalized to unit length along the feature dimension, whitened using the same teacher statistics, and normalized again, yielding $\widetilde{H}^S$ and $\widetilde{H}^T$. The loss is evaluated only at masked temporal positions $\mathcal{M}$ and averaged equally across the three layers:

$$\mathcal{L}_{\text{MCRP}} = \frac{1}{3|\mathcal{M}|} \sum_{\ell \in \{\text{Fusion,R3,R5}\}} \sum_{t \in \mathcal{M}} \left[1 - \cos\left(\widetilde{H}^S_{\ell,t}, \text{sg}(\widetilde{H}^T_{\ell,t})\right)\right].$$

Here, sg denotes stop-gradient; frequency masking perturbs the context. Whitening statistics are estimated from training examples without repeated weighting. The covariance is shrunk by 10% toward an identity matrix scaled by the mean variance, and its inverse square root defines the whitening matrix. The transform is fitted separately at each layer and then fixed. Pretraining learns the teacher's latent representations rather than reconstructing raw waveforms. After pretraining, the entire student is fine-tuned using sentence cross-entropy (CE):

$$\mathcal{L}_{\text{CE}} = -\log p_y.$$

During fine-tuning, the student undergoes sentence-supervised training with data augmentation. At deployment, complete AESSI uses only the student obtained after MCRP and CE fine-tuning to predict sentences.

### 3.6 Network Configuration

The time-frequency branch uses convolutional kernels of 5 × 5 and 3 × 5. Temporal kernel sizes in the frequency-aggregation and channel branches are 5 and 3. The Gram branch uses a 171-to-96 pointwise projection and a depthwise convolution with kernel width 5. The 96-to-96 linear gates at the outputs of the channel and Gram branches are initialized to zero and learned during training. The five backbone residual blocks use kernel width 5, dilation rates 1, 2, 4, 8, and 16, and dropout 0.18. The feedforward layer following attention expands to four times the feature width. The 160-dimensional projection head includes layer normalization, SiLU, and dropout.

Teacher training, MCRP, and student CE fine-tuning all use AdamW with learning rate $10^{-3}$, weight decay $10^{-4}$, batch size 32, and gradient-norm clipping at 5. MCRP runs for exactly 10 epochs. Teacher and student CE stages each run for at most 110 epochs. The learning rate is reduced by a factor of 0.5 after a plateau of 8 epochs, and early-stopping patience is 28 epochs.

### 3.7 Training and Test Splits

The main cross-day evaluation uses a personalized leave-one-day-out protocol, denoted LODO. For each of P002, P003, P004, P015, P018, and P024, the entire final session is held out as the target recording. Each fold contains 250 test trials and 43 source sessions. The target user's personal historical sessions precede their target date. Other users' source sessions are drawn from the available pool without further truncation at each target date. This is therefore a retrospective evaluation with an entire session held out.

For each historical session of the target user, the first eight blocks are used for training and the last two for validation. All 10 blocks of each source session from other users are used for training. Personal examples have sampling weight 8 and other-user examples weight 1. Mini-batches are sampled with replacement, with the number of draws per epoch equal to the natural number of training examples. This weighting emphasizes personal data. Validation accuracy is computed separately for each personal historical session and then averaged equally for model selection and early stopping. The target session is excluded from training, whitening-statistic estimation, and model selection; its labels are used for scoring only after predictions are fixed. Main results and component ablations use three random seeds (42, 123, and 456). We report each participant's mean decoding accuracy across these seeds.

### 3.8 Baseline Methods

We compare against TCN [4], EEGNet [22], ATCNet [2], Conv-Transformer [34], InoueNet [14], and SpeechNet [29]. InoueNet denotes the around-ear silent speech decoder proposed by Inoue et al., and SpeechNet is the decoding network used in SilentWear.

All methods start from the same 18-channel raw recordings and 3 s observation windows and use the same 25-class task, six complete target sessions, and personal-history training/validation splits. TCN, Xie's Conv-Transformer, EEGNet, and ATCNet receive the same full log-STFT as AESSI. TCN, EEGNet, and ATCNet only rearrange the electrode and frequency dimensions, retaining all frequency bins and time frames. SilentWear's SpeechNet and the Inoue decoder retain time-domain processing, receiving preprocessed signals at 500 Hz and 120 Hz, respectively, with their classification outputs adapted to 25 classes. All baselines use exactly the training and test splits described in Section 3.7, but neither augmentation nor MCRP. Appendix A provides implementation details.

### 3.9 Effects of Training Data

Beyond achieving cross-day online decoding without test-day calibration, practical deployment should reduce the data required from each user. We therefore examine whether other users' recordings can assist training and improve performance. To distinguish the roles of personal history and population data, we independently vary the number of personal historical sessions $H$ and other-user sessions $G$, evaluating all conditions on the same complete later-day recording. In particular, we examine accuracy when the user supplies zero, one, or two historical sessions.

This analysis includes P002, P003, P004, P015, P018, and P024, with $H = 0$, 1, or 2 and $G = 0, 4, 8, 12, 16, 20, 24, 28$, or all available population sessions. The condition $H = 0$ and $G = 0$ contains no training data and is omitted. When $H \geq 1$, we select the earliest $H$ personal sessions chronologically, using their first eight blocks for training and last two for validation. All 10 blocks of selected other-user sessions are used for training; personal and other-user examples have sampling weights of 8 and 1. Both data sources are used jointly in the same four-branch training pipeline. Models are selected by equally averaging validation accuracy over the selected personal historical sessions. Unselected personal sessions and the target test session are excluded from training, validation, teacher learning, whitening estimation, and model selection.

All data-volume conditions include these six participants and three random seeds (42, 123, and 456). Results are first averaged within participants and then equally across participants. All recordings of the target user are excluded from the other-user pool. A fixed sampling seed of 42 selects the sessions, and all three training seeds use the same selected data. The complete population pool contains 34 sessions for P003, 40 for P018, and 41 for each remaining target user.

When $H = 0$, all target-user recordings are excluded from training, validation, and statistical fitting. Each selected other-user session contributes its first eight blocks to training and last two to validation, with all example sampling weights equal to 1. Validation accuracy is averaged first across sessions within each source participant and then equally across source participants. The condition $H = 0$ and $G =$ full is strict leave-one-participant-out (LOPO) evaluation.

### 3.10 Transfer from a Shared to a Personal Sentence Set

To examine whether supervised experience with a limited shared sentence set helps users learn additional personal sentences, we compare personal-only training with transfer from population pretraining in the same six participants. The 25 sentences are divided by fixed original IDs into shared classes 01–10 and personally added classes 11–25, corresponding to zero-based labels 0–9 and 10–24, respectively (Table 3). All participants use the same split; classes are not selected according to test performance.

The population stage of the transfer condition uses only the 10 shared classes from other users. We first train a 10-class teacher and freeze it, perform 10 epochs of student MCRP, and then conduct sentence-supervised training on the 10 shared classes. Teacher training, whitening estimation, student MCRP, supervised training, and validation in this stage are all restricted to those shared classes. The first eight blocks of each selected population session are used for training and the last two for validation.

At the personal stage, the transfer condition loads all non-classifier parameters from the population student into a 25-class student, including the 160-dimensional pre-classification feature head and the MCRP projection layers, and reinitializes the 25-class linear classifier. The personal-only condition initializes a 25-class student with the same architecture from scratch. For each participant and random seed, a single personal teacher is trained using only the 25-class training data from that participant's entire set of historical sessions, and the corresponding whitening transform is fitted; both are shared by the two conditions. Both conditions then undergo 10 epochs of personal MCRP, after which the optimizer is reinitialized for sentence-supervised CE training with identical data augmentation for at most 110 epochs. Each personal session contributes its first eight blocks to training and its last two blocks to validation, with identical validation-based model selection and early-stopping rules. The 15 personally added classes have ground-truth labels during this stage, so the experiment evaluates the benefit of population pretraining for supervised personal learning rather than zero-shot recognition of new classes. Both conditions use the same personal teacher, whitening transform, personal training data, and training rules. They differ in how the personal student is initialized: the transfer condition inherits the population student's non-classifier parameters, whereas the personal-only condition starts from random initialization.

Table 3: The 10 shared classes used for population pretraining.

| Original sentence ID | Mandarin sentence and English translation |
|---|---|
| 01 | 我需要去厕所<br>I need to go to the restroom |
| 02 | 我想休息一会<br>I want to rest for a while |
| 03 | 帮我打个电话<br>Please make a phone call for me |
| 04 | 把我眼镜拿来<br>Bring me my glasses |
| 05 | 我想出去走走<br>I want to go out for a walk |
| 06 | 我需要换衣服<br>I need to change clothes |
| 07 | 请给我量体温<br>Please take my temperature |
| 08 | 请叫医生过来<br>Please call the doctor |
| 09 | 我渴了想喝水<br>I am thirsty and want some water |
| 10 | 我饿了想吃饭<br>I am hungry and want to eat |

IDs follow the acquisition program's original class mapping. Classes 01–10 are the 10 sentences for daily activity needs; the other 15 sentences belong to the social-interaction and physical/emotional-state groups in Table 1. Shared and personally added classes are evaluated using the same 25-class classifier within each model.

After personal MCRP and CE fine-tuning, both models are frozen and predict the same complete later-day recording. Each model uses one 25-class classifier. Accuracy is calculated by ground-truth label for the 10 shared classes, the 15 personally added classes, and all 25 classes. Because the added classes are labeled during personal training, this setting evaluates the benefit of population pretraining for supervised expansion of the sentence set.

### 3.11 Live Online Evaluation

Five of the six cross-day evaluation participants (P002, P003, P004, P015, and P018) returned and had their around-ear electrodes reapplied at least 21 days after their most recent historical recording. P024 was invited but could not return because of scheduling constraints. Each participant used a model trained in advance on all historical sessions, including the session previously held out for LODO, and frozen before the visit. Only random seed 42 was used. The remaining training procedure matched Section 3.7: the first eight blocks of each personal historical session were used for training and the last two for validation; all 10 blocks of each other-user source session were used for training. Personal and other-user examples had sampling weights of 8 and 1 and were sampled with replacement, with the number of draws per epoch equal to the natural number of training examples. Validation accuracy was computed for each personal historical session and then averaged equally for model selection and early stopping.

Each participant completed 50 live online trials, totaling 250 trials across five participants. Model weights, frontend parameters, and class mappings were fixed before the target session began. An experimenter applied the electrodes and checked signal quality for disconnected electrodes, defined as a channel remaining zero for 1 s; impedance was not rechecked. Channels 18 and 20 were checked for disconnection but not recorded, because the model had been trained with 18 rather than 20 channels. Four camera views (wide, right, opposite, and left) and screen capture recorded the evaluation continuously from electrode placement to completion (Figure 5). All five participants reviewed versions with personal information obscured and consented to their public release. Because of the 300 MB upload limit for Video Figure, only one representative evaluation video is provided with the paper. High-resolution recordings of all five participants will be made available after double-blind review.

For each trial, an independent tablet generated the target on site using the RANDOM.ORG integer generator (https://www.random.org/). The participant found the corresponding sentence on a printed copy of Table 1, memorized it, and pressed the space bar. During the green-cross cue, they fixated on the cross and silently articulated the sentence. After collecting 3 s of signals, the system applied its fixed preprocessing, predicted the sentence, and displayed the result. Participants then marked the result as correct by pressing 1 or incorrect by pressing 0. The authors checked these judgments against the recordings to verify their correctness and subsequently constructed a target–prediction table for each participant from the videos (supplementary materials). Online accuracy was calculated from the keypress logs and verified against the video-derived tables.

Computational performance was measured by trial-by-trial replay on the same CPU used for deployment. Each of the 250 raw online recordings was run once with its corresponding frozen model. Timing began with the raw observation window already in memory and ended when the sentence prediction was returned, including signal checks, log-STFT computation, the network forward pass, and prediction postprocessing. The platform was an Intel Core Ultra 9 185H with PyTorch 2.2.2 on CPU and batch size 1. Timing excluded signal acquisition, data loading, thread startup and waiting, log writing, and interface display.

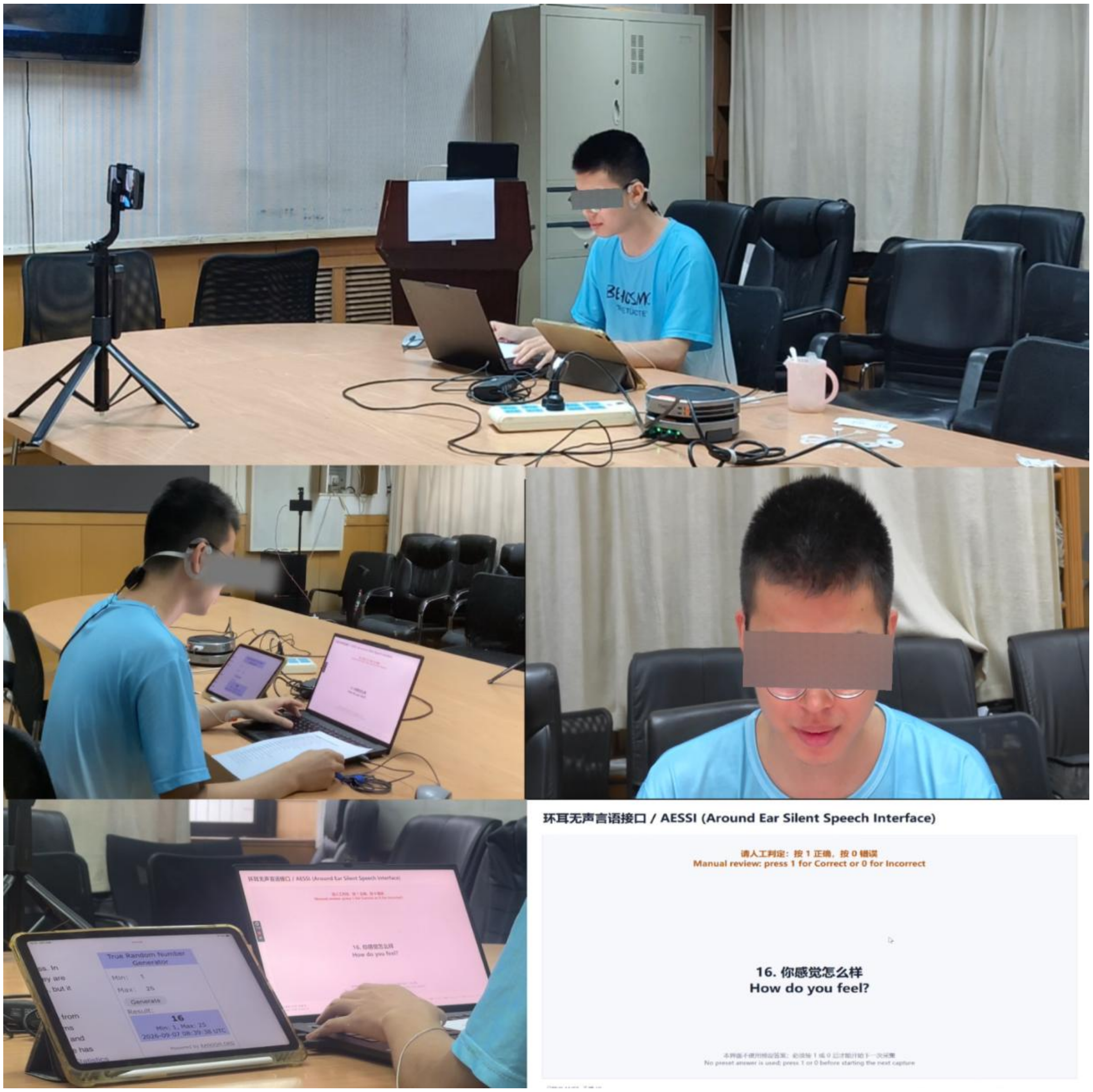

Figure 5: Example screenshots from the live online evaluation. After RANDOM.ORG generates 16, the participant finds sentence 16 ('How do you feel?') on the printed sentence list and silently articulates it. AESSI displays the corresponding prediction. Views show the participant, electrode placement, independent tablet, and system feedback; identifying facial information is obscured.

## 4 RESULTS

We first report decoding on complete later-day sessions and comparisons with baselines, followed by ablations of architecture and training strategies. We then present the effects of training-data volume, sentence-set transfer, and live online evaluation.

### 4.1 Cross-Day Decoding and Baseline Comparison

Table 4 presents the baseline results. Mean decoding accuracies were 20.80% for TCN, 17.20% for EEGNet, 41.67% for ATCNet, 21.13% for Conv-Transformer, 30.87% for InoueNet, and 35.60% for SpeechNet. By comparison, AESSI trained only with sentence supervision, without Aug or MCRP, achieved 74.16%; the complete training pipeline achieved 92.24%. Both exceeded the adapted baselines.

Table 4: Comparison of AESSI and adapted baseline models.

| Method | Decoding accuracy (%) |
|---|---|
| TCN [4] | 20.80 ± 23.87 |
| EEGNet [22] | 17.20 ± 6.22 |
| ATCNet [2] | 41.67 ± 12.59 |
| Conv-Transformer [34] | 21.13 ± 14.15 |
| InoueNet [14] | 30.87 ± 21.70 |
| SpeechNet [29] | 35.60 ± 18.33 |
| AESSI (w.o. Aug and MCRP) | 74.16 ± 12.62 |
| AESSI (ours) | 92.24 ± 3.81 |

Values are mean ± sample standard deviation across participants after averaging each participant's results over three random seeds. 'w.o.' means 'without.'

### 4.2 Roles of Decoder Branches and Training Strategies

Our multi-branch design covers several views of the signal with deliberate information overlap, allowing the remaining representations to support decoding when a branch is absent. Removing the time-frequency, frequency-aggregation, channel, or Gram branch and retraining produced mean accuracies of 91.67%, 91.67%, 91.82%, and 92.11%, respectively. The corresponding paired decreases relative to the complete model were 0.58, 0.58, 0.42, and 0.13 percentage points (Figure 6b). Across the evaluated configurations, removing any single branch changed accuracy only modestly, consistent with some substitutability among representations.

Figure 6c compares four combinations of the training pathway and supervised augmentation. Sentence-supervised AESSI without Aug or MCRP achieved 74.16% accuracy. Adding Aug increased accuracy to 77.20%, corresponding to direct classification by the teacher used in the MCRP pipeline. Deploying the student after MCRP and CE fine-tuning further increased accuracy to 92.24%. We also disabled Aug during both teacher training and student fine-tuning; even in this setting, MCRP increased accuracy from 74.16% to 89.67%. These results support the contribution of MCRP.

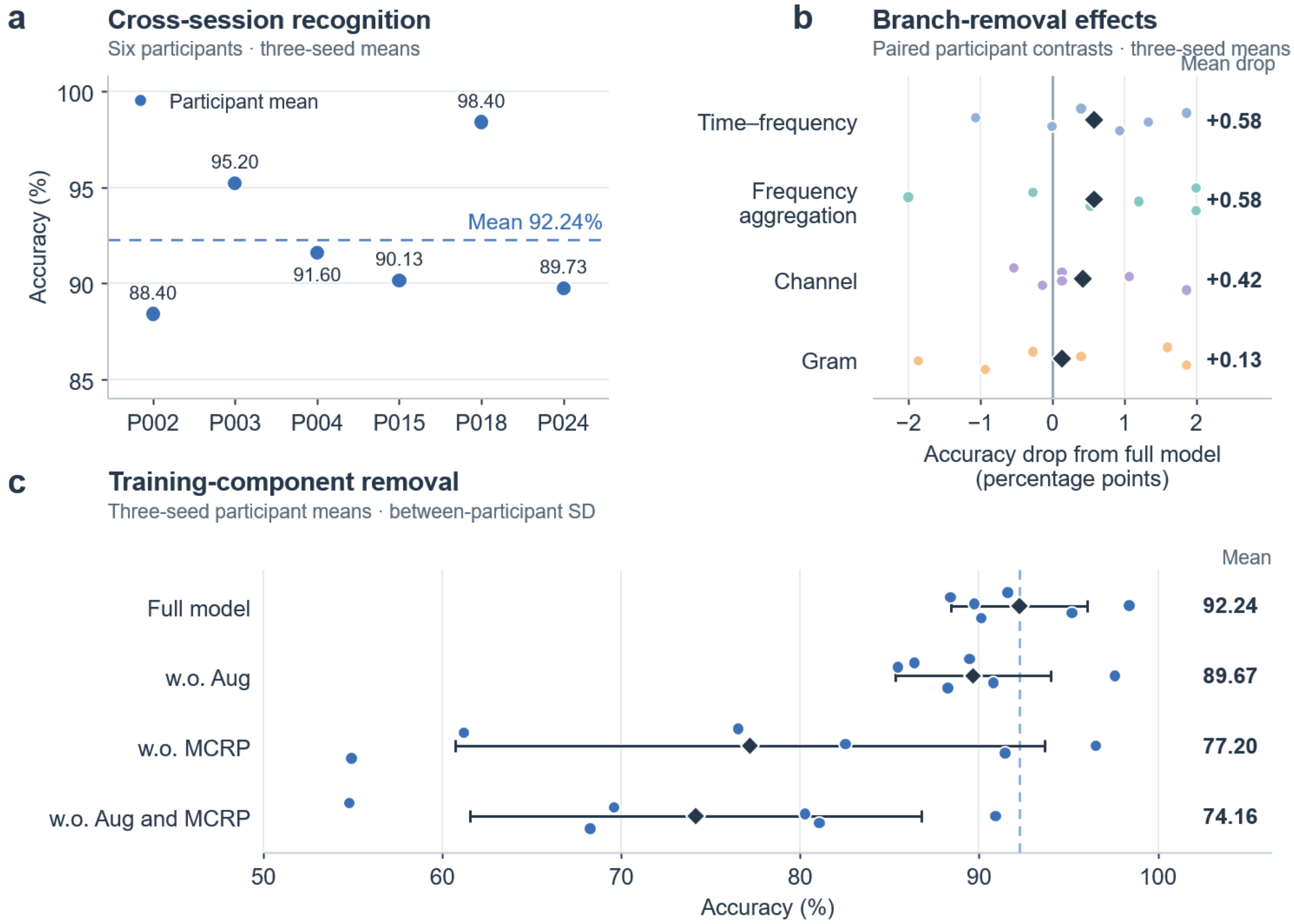

Figure 6: Cross-day results and component ablations for the four-branch model. (a) Results for the six participants using the complete model, averaged over three seeds within each participant. (b) Paired accuracy decreases after removing one branch, again averaged over three seeds per participant. A small decrease does not imply no contribution. (c) Effects of removing Aug, MCRP, or both; 'w.o.' means 'without.' Error bars show the sample standard deviation across participants' three-seed means.

### 4.3 Effects of Personal History and Population Training Data

Figure 7 reports the training-data-volume analysis for six participants. H = 0, $H = 1$, and $H = 2$ denote zero, one, and two personal historical sessions; $G$ is the number of other-user sessions. All conditions are evaluated on the same complete later-day session, examining the relationship between personal data-collection effort and use of population data.

The condition $H = 0$ and $G =$ full is strict LOPO evaluation: all target-user recordings are excluded from training, validation, and statistical fitting. The same four-branch pipeline achieved 68.44% mean accuracy across six participants and three seeds. At $H = 1$, increasing $G$ from 0 to the full population pool increased accuracy from 13.29% to 73.62%, a gain of 60.33 percentage points. At $H = 2$, accuracy increased from 35.07% to 88.58%, a gain of 53.51 points. Population data thus offered substantial benefits when personal history was limited. Adding a second personal session improved accuracy by 21.78 points at $G = 0$ and by 14.96 points with the full population pool, indicating the continued value of longitudinal personal recordings. Overall, accuracy tended to increase with G at each H, underscoring the value of additional other-user data for personalized decoding.

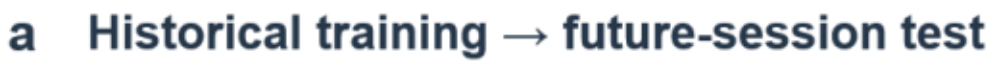

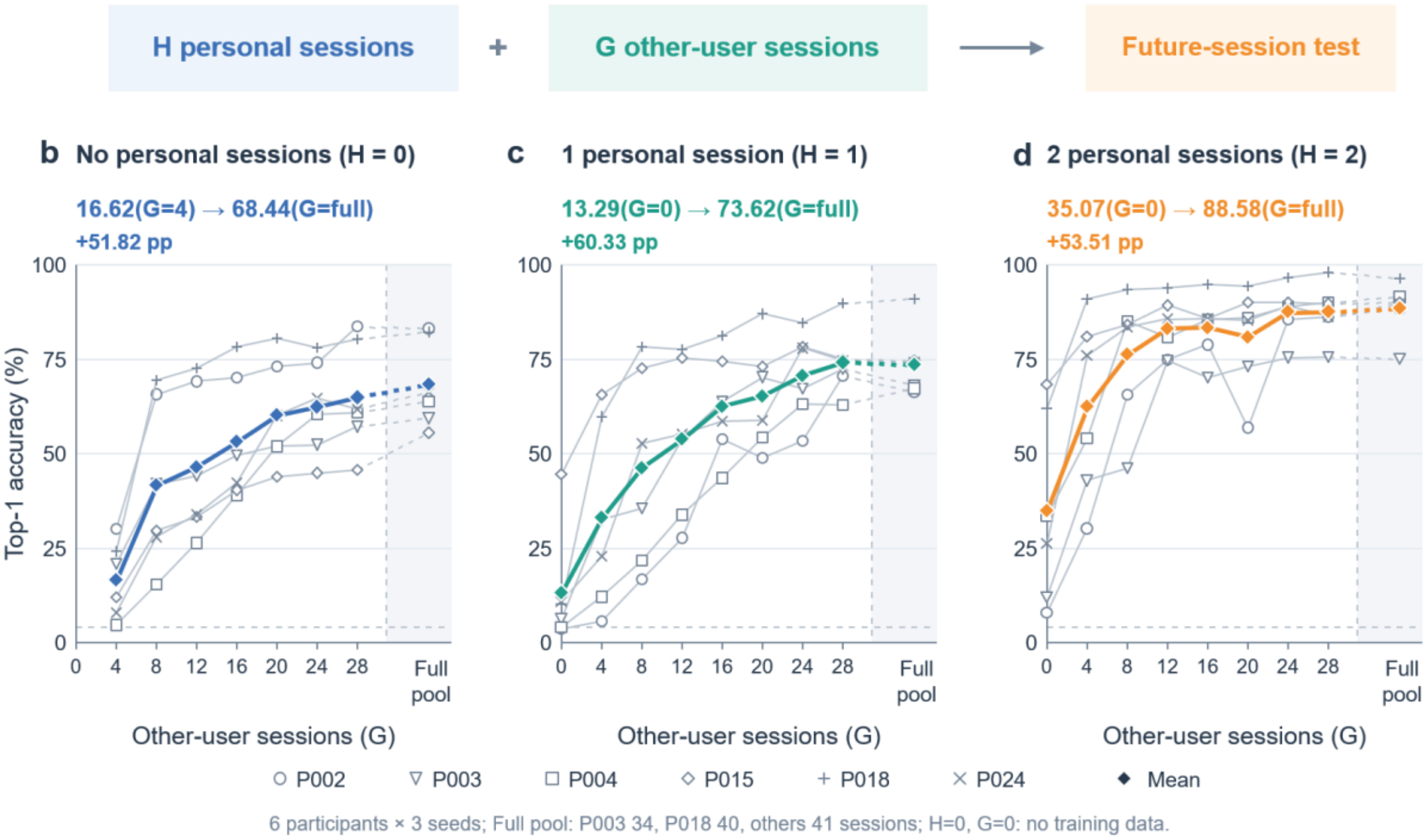

Figure 7: Effects of personal history and other-user data. (a) Models are trained on H personal historical sessions and G other-user sessions, frozen, and tested on a complete later-day session. (b–d) Results for six participants at H = 0, H = 1, and H = 2. Each participant's result is averaged over three seeds; gray markers show individuals and colored diamonds show the six-participant mean. At H = 0, increasing G from 4 to the full pool raises accuracy from 16.62% to 68.44%. At H = 1 and H = 2, increasing G from 0 to the full pool raises accuracy from 13.29% to 73.62% and from 35.07% to 88.58%, respectively. Numeric G values use proportional spacing. Full pools are shown separately because their sizes depend on the target user.

### 4.4 Transfer from Shared to Personal Sentence Sets

The sentence-set transfer experiment used the same six target participants. The transfer condition first completed population teacher training, student MCRP, and sentence-supervised training on other users' 10 shared classes, as described in Section 3.10. It then inherited the population student's non-classifier parameters and reinitialized a 25-class classifier. The personal-only condition initialized its student from scratch. Both used the same 25-class personal history, personal teacher, and whitening transform, and both completed personal MCRP and CE fine-tuning. After training, the models were frozen and evaluated on the same complete later-day recordings without test-day adaptation. Accuracies for the shared 10 classes, added 15 classes, and all 25 classes were calculated from each model's 25-class predictions.

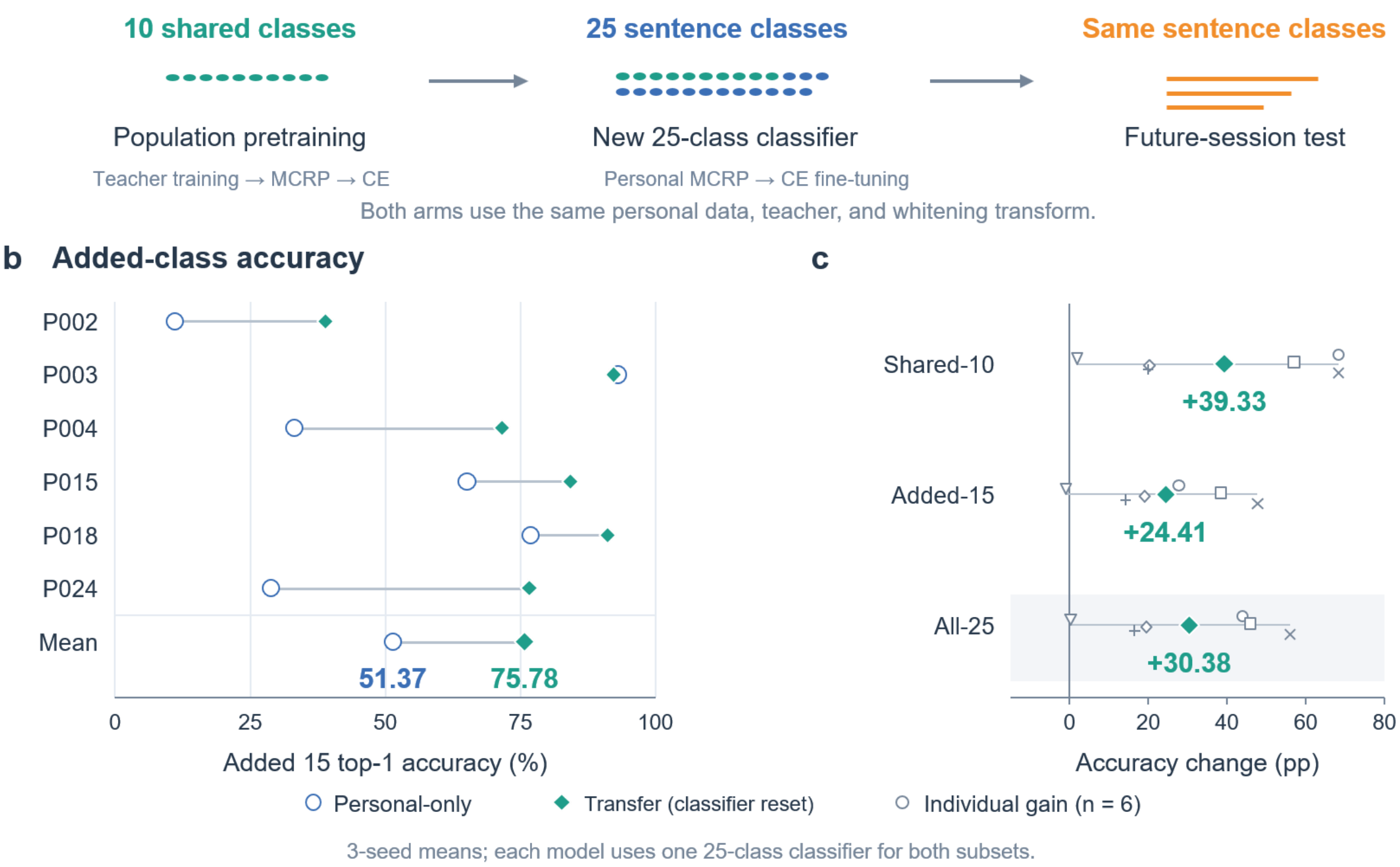

Figure 8: Transfer from a shared to a personal sentence set. (a) The transfer condition uses population pretraining on other users' 10 shared classes, inherits the population student's non-classifier parameters, and reinitializes a 25-class classifier. The personal-only student is randomly initialized. Both conditions use the same 25-class personal history, personal teacher, and whitening transform, complete personal MCRP and CE fine-tuning, and then freeze their models for testing on the full later-day session. (b) Paired results for the 15 added classes, averaged over three seeds within each participant. Mean accuracy is 51.37% for personal-only training and 75.78% for transfer. (c) Mean gains are 39.33, 24.41, and 30.38 percentage points for shared, added, and all classes, respectively. Negative individual gains are retained; gray segments show the observed range. Overall accuracy rises from 51.18% to 81.56%. Added classes have labels during personal training; all subsets are scored with each model's 25-class classifier.

With the same personal training procedure, initialization from population pretraining on the 10 shared classes increased accuracy relative to random student initialization from 50.89% to 90.22% for shared classes, from 51.37% to 75.78% for the 15 added classes, and from 51.18% to 81.56% overall. The gains were 39.33, 24.41, and 30.38 percentage points, respectively (Figure 8). Although the 15 added classes were excluded from population pretraining, they were labeled during personal training. These results support the benefit of population pretraining for subsequent supervised learning of additional personal sentences.

Transfer gains varied across individuals. After averaging over three seeds per participant, accuracy over all 25 classes improved for all six participants, by 0.27–56.00 percentage points. Accuracy on the 15 added classes improved for five participants but decreased by 0.89 points for P003 (Figure 8b,c). One possible explanation is that P003 had 10 sessions, nine of which were available for training. This amount of personal data may already have been sufficient to train an effective personalized system, reducing the value of other-user data.

### 4.5 Live Online Evaluation

The five participants had their electrodes reapplied after at least three weeks (21 days). An independent tablet generated targets; the live system acquired signals, predicted sentences with frozen models, and displayed the results (Figure 5). Electrode placement and signal-quality checks took approximately 5 min 26 s per participant on average, estimated from video: 4 min 40 s for P002, 6 min 40 s for P003, 4 min 58 s for P004, 5 min 40 s for P015, and 5 min 12 s for P018. AESSI required no calibration or adaptation, eliminating test-day calibration collection and adaptation waiting time. Use could begin as soon as signal checks were complete. Correct predictions out of 50 trials were 46 for P002, 50 for P003, 50 for P004, 50 for P015, and 49 for P018, corresponding to 92%, 100%, 100%, 100%, and 98% accuracy. In total, 245 of 250 trials were correct, yielding 98.00% online accuracy with a between-participant sample standard deviation of 3.46 percentage points.

Complete computational replay on the deployment CPU took a median of 31.70 ms and a 95th percentile of 35.1 ms. All 250 replay predictions matched the live records. This measure includes preprocessing and sentence prediction and characterizes computational overhead after the observation window is available. Live logs separately recorded a median network-forward-pass time of 24.30 ms.

## 5 DISCUSSION

### 5.1 Evaluating Reuse with Complete Sessions after Reapplication

AESSI evaluates decoding in the context of a user putting the device on again on a future day. Within-session training and test data share an electrode application, similar physiological states, the environment, and task timing (Figure 2a). High classification accuracy in this setting primarily establishes discriminative information within that recording. Holding out an entire later-day session instead preserves the joint changes in electrode position and contact, physiological state, and articulation strategy that occur during everyday reuse. Our additional online evaluation demonstrates cross-day online decoding without test-day calibration.

### 5.2 Multiple Representational Views and MCRP

The four branches characterize the same observation window through local time-frequency patterns, spectral trajectories, channel activity trajectories, and similarity between channels' spectral shapes. Removing any one branch reduced mean accuracy by 0.13–0.58 percentage points. Thus, the complete model achieved high mean accuracy across these configurations, but the marginal gain from any single branch was limited. We do not claim that any individual branch is particularly important. Instead, the experimental results show that removing any branch does not lead to a substantial decrease in performance, indicating a degree of redundancy and substitutability in the system design.

Figure 6c shows that the MCRP student outperformed direct teacher classification both with and without supervised augmentation: 92.24% versus 77.20% with Aug, and 89.67% versus 74.16% without Aug. The directly CE-trained teacher and final student use the same four-branch decoding architecture. This comparison therefore evaluates a training pathway that adds masked teacher-representation pretraining followed by student CE fine-tuning. Together, the four conditions support adding MCRP to CE-supervised learning, with complete AESSI achieving its highest mean accuracy when combined with Aug. These results support the contribution of MCRP.

### 5.3 Differences from Baseline Models

AESSI substantially outperformed the adapted baselines. The complete training strategy exceeded the strongest baseline by 50.57 percentage points; sentence-supervised AESSI without Aug or MCRP exceeded it by 32.49 points. These gaps may partly reflect the original models' intended tasks and the adaptations required here. For example, Inoue et al.'s system [14] was designed for word classification and collected personal calibration data for fine-tuning before online testing. Our setting prohibits test-day fine-tuning. SpeechNet in SilentWear [29] was designed for low-power neck-EMG command recognition, with offline cross-session evaluation and on-device deployment measurements. Its offline preprocessing applied bidirectional zero-phase filtering to continuous recordings before segmentation. To avoid using signals after the observation window, we restricted filtering to the acquired 3 s window, changing the filter boundary conditions. These models may be better suited to the conditions examined in their original studies. Applying them directly to cross-day online decoding without test-day calibration may therefore yield lower performance in our setting.

### 5.4 Personal History and Population Data

The six-participant H/G experiment shows that additional other-user data improve later-day recognition when personal history is limited (Figure 7). At $H = 0$, with no personal history, increasing G from 4 to the full population pool raised mean accuracy from 16.62% to 68.44% using the four-branch configuration. At H = 1, increasing G from 0 to the full pool raised accuracy from 13.29% to 73.62%; at H = 2, it rose from 35.07% to 88.58%. The respective gains from population data were 51.82, 60.33, and 53.51 percentage points. These results support jointly using population data and longitudinal personal recordings to accommodate a user's articulation patterns and measurement conditions.

### 5.5 Shared and Personal Sentence Sets

Figure 8 further examines whether population pretraining on a shared sentence set helps users learn additional personal classes. The transfer condition inherits the non-classifier parameters of a student trained on other users' 10 shared classes, whereas the control randomly initializes its personal student. Both then use the same 25-class personal history, personal teacher, and whitening transform and complete identical personal MCRP and CE fine-tuning. Under these conditions, mean accuracy on the 15 added classes increased from 51.37% to 75.78%, and overall accuracy increased from 51.18% to 81.56%, gains of 24.41 and 30.38 percentage points. The added classes were excluded from population pretraining, and the transfer condition's 25-class classifier was reinitialized. These findings therefore support the benefit of population-pretrained parameters for supervised learning of additional personal classes and motivate exploring larger candidate sentence sets using existing data.

### 5.6 From Fixed-Window Prediction to Online Interaction

Live recordings from five participants show that AESSI can process newly arriving signals after electrode reapplication and return a sentence prediction after a fixed articulation window. Targets generated on an independent device, continuous video from electrode placement, and per-trial records jointly support verification of preparation, participant actions, and system outputs. Retrospective full-session holdout and prospective live follow-up together support cross-day reuse of historical models. The prospective evaluation additionally tests the integration of acquisition, preprocessing, prediction, and feedback.

The current system makes one classification after acquiring a 3 s observation window. Replay of 250 raw live recordings on the deployment CPU yielded a median preprocessing and sentence-prediction time of 31.70 ms, indicating modest computational overhead. The videos also show that the system provides results almost immediately after acquisition ends, and all five participants reported no noticeable delay.

### 5.7 Limitations and Future Work

AESSI currently recognizes 25 predefined short Mandarin sentences and does not support open-vocabulary or freely composed sentence input. Although the sentences reflect everyday communication needs, performance has been validated only in healthy participants. Applicability to intended users, including people with post-stroke dysarthria, requires further study. Future work will expand the candidate sentence set, examine how additional personal recordings and pretraining data affect accuracy and collection effort, and include participants with post-stroke dysarthria.

The prototype uses the space bar to initiate a 3 s acquisition window and 1/0 keypresses after prediction to label correctness during evaluation. Practical use will require further investigation of initiation, confirmation, and error-correction mechanisms that reduce dependence on a keyboard. Blinks or eye movements are possible control signals, but their coordination with silent-articulation recognition, false activations, and interaction burden need dedicated evaluation.

Around-ear recordings contain multiple electrophysiological components. Articulation-related facial and jaw EMG or EEG in these recordings can provide useful information for silent speech recognition. This study focuses on the decodability and cross-day reuse of mixed around-ear signals during silent articulation; it does not separately quantify the contributions of different physiological sources. Participants were instructed to fixate on a green cross during the 3 s observation window to reduce variation in eye movements, but this measure alone cannot rule out effects of non-articulatory movements. Targeted behavioral controls and synchronized recordings could further examine sensitivity to these sources of variation.

## 6 CONCLUSION

We present AESSI for reusing silent speech models after a user reapplies a device on a future day. AESSI combines around-ear log-STFT inputs, four-branch representations, and single-teacher masked-context representation pretraining. In a dataset of 44 sessions from 24 participants, classification accuracy on six complete later-day target sessions averaged 92.24% across three random seeds. Branch-removal results are consistent with partially substitutable information across paths. Four comparisons of training strategies further show that, with or without supervised augmentation, the student after MCRP achieves higher mean decoding accuracy than the directly supervised teacher, supporting the empirical value of this training pathway in the evaluated setting.

The training-data-volume and shared-sentence-set transfer results support using population experience to help establish personal models while retaining the value of longitudinal personal recordings. Prospective online evaluation with five participants after at least 21 days achieved 98.00% accuracy, providing further evidence that the frozen four-branch pipeline can operate directly after electrode reapplication. Future work can examine stability across more users, repeated follow-ups, and naturalistic use, and develop natural initiation and error-correction mechanisms.

## REFERENCES

[1] Ali Abdolrahmani, Maya Howes Gupta, Mei-Lian Vader, Ravi Kuber, and Stacy Branham. 2021. Towards more transactional voice assistants: Investigating the potential for a multimodal voice-activated indoor navigation assistant for blind and sighted travelers. In *Proceedings of the 2021 CHI Conference on Human Factors in Computing Systems*. Association for Computing Machinery, New York, NY, USA, Article 495, 16 pages. https://doi.org/10.1145/3411764.3445638

[2] Hamdi Altaheri, Ghulam Muhammad, and Mansour Alsulaiman. 2023. Physics-informed attention temporal convolutional network for EEG-based motor imagery classification. *IEEE Transactions on Industrial Informatics* 19, 2 (2023), 2249–2258. https://doi.org/10.1109/TII.2022.3197419

[3] Alexei Baevski, Wei-Ning Hsu, Qiantong Xu, Arun Babu, Jiatao Gu, and Michael Auli. 2022. data2vec: A general framework for self-supervised learning in speech, vision and language. In *Proceedings of the 39th International Conference on Machine Learning*, Vol. 162. PMLR, 1298–1312. https://proceedings.mlr.press/v162/baevski22a.html

[4] Shaojie Bai, J. Zico Kolter, and Vladlen Koltun. 2018. An empirical evaluation of generic convolutional and recurrent networks for sequence modeling. arXiv:1803.01271. https://arxiv.org/abs/1803.01271

[5] Martin G. Bleichner and Stefan Debener. 2017. Concealed, unobtrusive ear-centered EEG acquisition: cEEGrids for transparent EEG. *Frontiers in Human Neuroscience* 11, Article 163 (2017). https://doi.org/10.3389/fnhum.2017.00163

[6] Stefan Debener, Reiner Emkes, Maarten De Vos, and Martin Bleichner. 2015. Unobtrusive ambulatory EEG using a smartphone and flexible printed electrodes around the ear. *Scientific Reports* 5, 1, Article 16743 (2015). https://doi.org/10.1038/srep16743

[7] Florian Denk, Marleen Grzybowski, Stephan M. A. Ernst, Birger Kollmeier, Stefan Debener, and Martin G. Bleichner. 2018. Event-related potentials measured from in and around the ear electrodes integrated in a live hearing device for monitoring sound perception. *Trends in Hearing* 22, Article 2331216518788219 (2018). https://doi.org/10.1177/2331216518788219

[8] Xuefu Dong, Yifei Chen, Yuuki Nishiyama, Kaoru Sezaki, Yuntao Wang, Ken Christofferson, and Alex Mariakakis. 2024. ReHEarSSE: Recognizing hidden-in-the-ear silently spelled expressions. In *Proceedings of the CHI Conference on Human Factors in Computing Systems*. Association for Computing Machinery, New York, NY, USA, 1–16. https://doi.org/10.1145/3613904.3642095

[9] Margaret Foley, Géry Casiez, and Daniel Vogel. 2020. Comparing smartphone speech recognition and touchscreen typing for composition and transcription. In *Proceedings of the 2020 CHI Conference on Human Factors in Computing Systems*. Association for Computing Machinery, New York, NY, USA, 1–11. https://doi.org/10.1145/3313831.3376861

[10] David Gaddy. 2022. *Voicing silent speech.* Ph.D. Dissertation. University of California, Berkeley. https://escholarship.org/uc/item/2gh5j6g7

[11] David Gaddy and Dan Klein. 2020. Digital voicing of silent speech. In *Proceedings of the 2020 Conference on Empirical Methods in Natural Language Processing (EMNLP)*. Association for Computational Linguistics, 5521–5530. https://doi.org/10.18653/v1/2020.emnlp-main.445

[12] David Gaddy and Dan Klein. 2021. An improved model for voicing silent speech. In *Proceedings of the 59th Annual Meeting of the Association for Computational Linguistics and the 11th International Joint Conference on Natural Language Processing (Volume 2: Short Papers)*. Association for Computational Linguistics, 175–181. https://doi.org/10.18653/v1/2021.acl-short.23

[13] Kaiming He, Xinlei Chen, Saining Xie, Yanghao Li, Piotr Dollár, and Ross Girshick. 2022. Masked autoencoders are scalable vision learners. In *Proceedings of the IEEE/CVF Conference on Computer Vision and Pattern Recognition*. 16000–16009. https://openaccess.thecvf.com/content/CVPR2022/html/He_Masked_Autoencoders_Are_Scalable_Vision_Learners_CVPR_2022_paper.html

[14] Masakazu Inoue, Eri Hatakeyama, Yuya Kita, and Shuntaro Sasai. 2026. Large-scale training data enhances silent speech decoding with around-ear EEG. *Journal of Neural Engineering* 23, 2, Article 026027 (2026). https://doi.org/10.1088/1741-2552/ae54d0

[15] Masakazu Inoue, Motoshige Sato, Kenichi Tomeoka, Nathania Nah, Eri Hatakeyama, Kai Arulkumaran, Ilya Horiguchi, and Shuntaro Sasai. 2025. A silent speech decoding system from EEG and EMG with heterogenous electrode configurations. In *Interspeech 2025*. ISCA, 5603–5607. https://doi.org/10.21437/Interspeech.2025-1183

[16] Muyun Jiang, Wei Zhang, Yi Ding, Kok Ann Colin Teo, LaiGuan Fong, Shuailei Zhang, Zhiwei Guo, Chenyu Liu, Raghavan Bhuvanakantham, Wei Khang Jeremy Sim, Chuan Huat Vince Foo, Rong Hui Jonathan Chua, Parasuraman Padmanabhan, Victoria Leong, Jia Lu, Balázs Gulyás, and Cuntai Guan. 2026. Decoding covert speech from EEG by functional areas spatio-temporal transformer. *IEEE Journal of Biomedical and Health Informatics* 30, 6 (2026), 4974–4984. https://doi.org/10.1109/JBHI.2026.3653025

[17] Simon L. Kappel, David Looney, Danilo P. Mandic, and Preben Kidmose. 2017. Physiological artifacts in scalp EEG and ear-EEG. *BioMedical Engineering OnLine* 16, 1, Article 103 (2017). https://doi.org/10.1186/s12938-017-0391-2

[18] Arnav Kapur, Shreyas Kapur, and Pattie Maes. 2018. AlterEgo: A personalized wearable silent speech interface. In *Proceedings of the 23rd International Conference on Intelligent User Interfaces*. Association for Computing Machinery, New York, NY, USA, 43–53. https://doi.org/10.1145/3172944.3172977

[19] Naoki Kimura, Tan Gemicioglu, Jonathan Womack, Richard Li, Yuhui Zhao, Abdelkareem Bedri, Zixiong Su, Alex Olwal, Jun Rekimoto, and Thad Starner. 2022. SilentSpeller: Towards mobile, hands-free, silent speech text entry using electropalatography. In *CHI Conference on Human Factors in Computing Systems*. Association for Computing Machinery, New York, NY, USA, Article 288, 19 pages. https://doi.org/10.1145/3491102.3502015

[20] Naoki Kimura, Michinari Kono, and Jun Rekimoto. 2019. SottoVoce: An ultrasound imaging-based silent speech interaction using deep neural networks. In *Proceedings of the 2019 CHI Conference on Human Factors in Computing Systems*. Association for Computing Machinery, New York, NY, USA, 1–11. https://doi.org/10.1145/3290605.3300376

[21] Michael T. Knierim, Max Schemmer, and Monica Perusquía-Hernández. 2021. Exploring the recognition of facial activities through around-the-ear electrode arrays (cEEGrids). In *Information systems and neuroscience: NeuroIS Retreat 2021*. Springer, 47–55. https://doi.org/10.1007/978-3-030-88900-5_6

[22] Vernon J Lawhern, Amelia J Solon, Nicholas R Waytowich, Stephen M Gordon, Chou P Hung, and Brent J Lance. 2018. EEGNet: A compact convolutional neural network for EEG-based brain–computer interfaces. *Journal of Neural Engineering* 15, 5, Article 056013 (2018). https://doi.org/10.1088/1741-2552/aace8c

[23] Ren Li, Jared S. Johansen, Hamad Ahmed, Thomas V. Ilyevsky, Ronnie B. Wilbur, Hari M. Bharadwaj, and Jeffrey Mark Siskind. 2021. The perils and pitfalls of block design for EEG classification experiments. *IEEE Transactions on Pattern Analysis and Machine Intelligence* 43, 1 (2021), 316–333. https://doi.org/10.1109/TPAMI.2020.2973153

[24] Zhao Li, Bin Ma, Weifan Mao, Jianxing Zhang, Zhuting Yu, and Yizhou Lu. 2024. SVIT-SSR: A sEMG-based vision transformer approach for silent speech recognition. *Electronics Letters* 60, 21, Article e13285 (2024). https://doi.org/10.1049/ell2.13285

[25] Geoffrey S. Meltzner, James T. Heaton, Yunbin Deng, Gianluca De Luca, Serge H. Roy, and Joshua C. Kline. 2017. Silent speech recognition as an alternative communication device for persons with laryngectomy. *IEEE/ACM Transactions on Audio, Speech, and Language Processing* 25, 12 (2017), 2386–2398. https://doi.org/10.1109/TASLP.2017.2740000

[26] Bojana Mirkovic, Martin G. Bleichner, Maarten De Vos, and Stefan Debener. 2016. Target speaker detection with concealed EEG around the ear. *Frontiers in Neuroscience* 10, Article 349 (2016). https://doi.org/10.3389/fnins.2016.00349

[27] David A. Moses, Sean L. Metzger, Jessie R. Liu, Gopala K. Anumanchipalli, Joseph G. Makin, Pengfei F. Sun, Josh Chartier, Maximilian E. Dougherty, Patricia M. Liu, Gary M. Abrams, Adelyn Tu-Chan, Karunesh Ganguly, and Edward F. Chang. 2021. Neuroprosthesis for decoding speech in a paralyzed person with anarthria. *New England Journal of Medicine* 385, 3 (2021), 217–227. https://doi.org/10.1056/NEJMoa2027540

[28] Rui Song, Xu Zhang, Xi Chen, Xiang Chen, Xun Chen, Shuang Yang, and Erwei Yin. 2023. Decoding silent speech from high-density surface electromyographic data using transformer. *Biomedical Signal Processing and Control* 80, Article 104298 (2023). https://doi.org/10.1016/j.bspc.2022.104298

[29] Giusy Spacone, Sebastian Frey, Giovanni Pollo, Alessio Burrello, Daniele Jahier Pagliari, Victor Kartsch, Andrea Cossettini, and Luca Benini. 2026. SilentWear: An ultra-low power wearable system for EMG-based silent speech recognition. arXiv:2603.02847. https://arxiv.org/abs/2603.02847

[30] Chenyu Tang, Shuo Gao, Cong Li, Wentian Yi, Yuxuan Jin, Xiaoxue Zhai, Sixuan Lei, Hongbei Meng, Zibo Zhang, Muzi Xu, Shengbo Wang, Xuhang Chen, Chenxi Wang, Hongyun Yang, Ningli Wang, Wenyu Wang, Jin Cao, Xiaodong Feng, Peter Smielewski, Yu Pan, Wenhui Song, Martin Birchall, and Luigi G. Occhipinti. 2026. Wearable intelligent throat enables natural speech in stroke patients with dysarthria. *Nature Communications* 17, 1, Article 293 (2026). https://doi.org/10.1038/s41467-025-68228-9

[31] Chenyu Tang, Josée Mallah, Dominika Kazieczko, Wentian Yi, Tharun Reddy Kandukuri, Edoardo Occhipinti, Bhaskar Mishra, Sunita Mehta, and Luigi G. Occhipinti. 2025. Wireless silent speech interface using multichannel textile EMG sensors integrated into headphones. *IEEE Transactions on Instrumentation and Measurement* 74, Article 4013710 (2025). https://doi.org/10.1109/TIM.2025.3583386

[32] Chenyu Tang, Liang Qi, Shuo Gao, Zibo Zhang, Wentian Yi, Muzi Xu, Edoardo Occhipinti, Yu Pan, and Luigi G. Occhipinti. 2026. Sensing technologies for silent speech interfaces. *Nature Sensors* 1, 1 (2026), 16–26. https://doi.org/10.1038/s44460-025-00010-2

[33] Xue Wang, Zixiong Su, Jun Rekimoto, and Yang Zhang. 2024. Watch your mouth: Silent speech recognition with depth sensing. In *Proceedings of the CHI Conference on Human Factors in Computing Systems*. Association for Computing Machinery, New York, NY, USA, 1–15. https://doi.org/10.1145/3613904.3642092

[34] Liang Xie, Yakun Zhang, Hao Yuan, Meishan Zhang, Xingyu Zhang, Changyan Zheng, Ye Yan, and Erwei Yin. 2025. Neural Chinese silent speech recognition with facial electromyography. *Speech Communication* 171, Article 103230 (2025). https://doi.org/10.1016/j.specom.2025.103230

[35] Xiran Xu, Bo Wang, Boda Xiao, Yadong Niu, Yiwen Wang, Xihong Wu, Heping Cheng, and Jing Chen. 2026. The impacts of temporal autocorrelations on EEG decoding. *Biomedical Signal Processing and Control* 113, Article 108783 (2026). https://doi.org/10.1016/j.bspc.2025.108783

[36] Chaoqi Yang, M. Brandon Westover, and Jimeng Sun. 2023. ManyDG: Many-domain generalization for healthcare applications. In *The Eleventh International Conference on Learning Representations*. https://openreview.net/forum?id=lcSfirnflpW

[37] Jinzhao Zhou, Daniel Leong, Zehong Cao, Thomas Do, Sheng-Fu Liang, Tzyy-Ping Jung, and Chin-Teng Lin. 2025. MindSpeak: A real-time BCI system for silent speech. In *Proceedings of the 33rd ACM International Conference on Multimedia*. Association for Computing Machinery, New York, NY, USA, 13549–13551. https://doi.org/10.1145/3746027.3754486

## A BASELINE METHODS AND ADAPTATION DETAILS

The common supervised stage for all methods uses AdamW with initial learning rate $10^{-3}$, weight decay $10^{-4}$, batch size 32, gradient-norm clipping at 5, and at most 110 epochs. The learning rate is reduced by a factor of 0.5 after 8 plateau epochs; early-stopping patience is 28 epochs. Checkpoints are selected by equally averaging accuracy across personal historical validation sessions. These settings are fixed for this comparison, with no additional search over normalization or initial learning rate.

TCN. TCN [4] combines the electrode and frequency dimensions of the 18 × 100 × 61 log-STFT into a sequence with 1,800 feature channels and 61 time frames. The network contains four residual temporal convolutional blocks with 128 output channels each. Each block contains two 1D dilated causal convolutions with kernel length 3. Dilation rates are 1, 2, 4, and 8 across the four blocks, with dropout 0.2. Temporal mean pooling and a linear classifier produce the 25-class prediction.

EEGNet and ATCNet. Both rearrange the complete 18 × 100 × 61 log-STFT into 1,800 × 61 in a fixed electrode-major, frequency-minor order. Each electrode–frequency pair becomes an input feature channel, and all 61 time frames are retained. EEGNet extracts features through temporal convolution, depthwise convolution across input channels, and separable convolution. ATCNet forms three overlapping temporal windows over convolutional features, applies self-attention and temporal convolution within each, and averages their classification scores [2, 22]. To accommodate 61 input frames, both use F1 = 8 temporal filters, depth multiplier D = 2, and temporal kernel length 15. EEGNet uses separable-convolution kernel length 8; ATCNet has model width 16. Both retain the implementation's max-norm weight constraints, including a bound of 0.25 on EEGNet's classifier.

Conv-Transformer. Conv-Transformer [34] retains the original encoder's two-layer 2D convolutional subsampling, positional encoding, and six Transformer layers, with hidden dimension 256, eight attention heads, and feedforward width 512. We replace the original eight-electrode, 36-band input with the complete 18 × 100 × 61 log-STFT and adjust the input convolutional channels and projection after subsampling accordingly. The 2D convolutions extract temporal and frequency features, followed by Transformer modeling of temporal dependencies. Temporal mean pooling and a 25-class head perform sentence classification, replacing the original character-sequence CTC readout.

SpeechNet. Our adaptation retains the five-convolution-block SpeechNet backbone from SilentWear [29]. The first three blocks extract and pool temporal features, while the final two fuse cross-electrode features; global average pooling and a classifier produce predictions. We adjust the number of input electrodes to 18 and the number of output classes to 25. Each 3 s observation window is demeaned channelwise and resampled to 500 Hz, yielding an 18 × 1,500 time-domain input. Preprocessing uses the original study's 20 Hz high-pass and 50 Hz notch filters, but filtering is applied after window segmentation rather than before it.

InoueNet. We implement the classification branch described by Inoue et al. [14]: temporal convolutions extract features, a Hilbert transform obtains amplitude envelopes, and cross-electrode feature fusion and a Conformer encoder produce a classification-token output for 25-class prediction. The adaptation does not include the original system's audio embeddings, auxiliary mora-sequence supervision, or large-scale data pretraining. Preprocessing is confined to the already extracted 3 s window. First, thresholds estimated only from training data identify and zero abnormal channels. We then apply 27 Hz and 50 Hz notch filters, common-average referencing over valid channels, a 1–60 Hz band-pass filter, and resampling to 120 Hz. Finally, each channel is z-score normalized within the window, and samples with absolute z-scores above 3 are zeroed, yielding an 18 × 360 input.